\documentclass[12pt,preprint]{emulateapj}
\usepackage{longtable}

\shorttitle{Unexpected series of regular frequency spacing of Delta Scuti stars }
\shortauthors{Papar\'o et al.}

\begin{document}

\title{Unexpected series of regular frequency spacing of $\delta$ Scuti stars in the non-asymptotic regime -- I. The methodology
}

\author{M. Papar\'o\altaffilmark{1}, J.~M. Benk\H{o}\altaffilmark{1}, M. Hareter,\altaffilmark{1} and J.A. Guzik\altaffilmark{2}} 

\altaffiltext{1}{Konkoly Observatory, MTA CSFK, Konkoly Thege Mikl\'os \'ut 15-17., H-1121 Budapest, Hungary}
\altaffiltext{2}{Los Alamos National Laboratory, Los Alamos NM 87545 USA}

\email{paparo@konkoly.hu}

\begin{abstract}

A  sequence search method was developed to search regular frequency spacing in $\delta$ Scuti stars 
by visual inspection and algorithmic search. We searched for sequences of quasi-equally spaced 
frequencies, containing at least four members per sequence, in 90 $\delta$ Scuti stars observed by CoRoT. 
We found an unexpectedly large number of independent series of regular frequency spacing in 77 $\delta$ Scuti 
stars (from 1 to 8 sequences) in the non-asymptotic regime. 
We introduce the sequence search method presenting the sequences and echelle diagram of CoRoT 102675756 and 
the structure of the algorithmic search. Four sequences (echelle ridges) were found in the 5-21 d$^{-1}$ region, 
where the pairs of the sequences are shifted (between 0.5-0.59 d$^{-1}$) by twice the value of the estimated rotational splitting frequency (0.269 d$^{-1}$). 
%In the ray dynamic approach the explanation could be two sequences with two different $\hat{l}$ values and with 
%different parities. 
The general conclusions for the whole sample are also presented in this paper. 
The statistics of the spacings derived by the sequence search method, by FT and that of the shifts are also compared. 
In many stars, more than one almost equally valid spacing appeared. 
The model frequencies of FG Vir and their rotationally split components were used to reveal 
a possible explanation that one spacing is the large separation, while the other is a sum of the large 
separation and the rotational frequency.  In CoRoT 102675756, the two spacings (2.249 and 1.977 d$^{-1}$) agree better with the sum of a possible 1.710 d$^{-1}$ large separation and two or one times, respectively, the value of the rotational frequency.

\end{abstract}


\section{Introduction}

Delta Scuti stars could be very important targets of asteroseismology once 
mode identification is successfully performed. They lie on and above the main 
sequence with intermediate mass and spectral types between A2 and F5. Both 
radial and non-radial p-type and g-type modes are excited covering a wide 
range of frequencies between 5-50 d$^{-1}$, or even wider 
that was revealed in {\it Kepler} data by \citet{Balona11}. 
The appearance of the convective core introduces poorly known physical 
processes in the stellar interiors, such as convective overshoot, mixing of 
chemical elements and redistribution of angular momentum \citep{Zahn92}. The 
investigation of the latter processes has nowadays become an observational science 
(\citealt{Kurtz14, Saio15}). Both in KIC 1145123 and KIC 9244992 slightly different 
surface-to-core rotation velocities were found using g-mode triplets and even p-mode triplets and multiplets. 
The successful investigation of the $\delta$ Sct/$\gamma$ Dor hybrids relies on the very 
slow rotation of these stars. The equatorial rotation velocity is found to be about 
1~kms$^{-1}$ and the mean rotational splitting is 0.0138~d$^{-1}$.

However, most of the $\delta$ Scuti stars are intermediate or fast rotators 
with $\approx$ 100-200 kms$^{-1}$ equatorial rotational velocity. Due to the complexity of 
the oscillation spectra, their pulsation behavior is not fully understood, especially 
the rotation-pulsation interaction. The problems and prospects were reviewed before 
the CoRoT and {\it Kepler} space missions by \citet{Goupil05}. 
Mostly the perturbative theory was used in the interpretations. Despite the problems, in 
the last 20 years several attempts have been made to interpret the observed spectra 
of $\delta$ Scuti stars (see references in \citealt{Fox06}).

\citet{Lignieres06} showed that the rotation-pulsation interaction cannot be described using 
the perturbation theory for rapidly rotating stars. At the same time \citet{Roxburgh06} presented 
self-consistent two-dimensional models of main-sequence stars. \citet{Reese06} and \citet{Suarez10} 
pointed out that the most severe problems appear in stars that show very high surface velocities, 
such as $\delta$ Scuti and Be stars, or in stars where the surface rotates slowly, but in which the 
pulsation periods are of the same order as the rotation period, such as for SPB and $\gamma$ Doradus stars.

After solving a basic convergence problem (\citealt{Jackson04, McGregor07}), the modeling 
of rapidly rotating stars greatly improved. The models revealed four families of modes 
(low frequency modes, whispering gallery modes, chaotic modes and island modes). The frequency spectrum 
of rotating stars is interpreted as the superposition of subspectra of different mode families. 
The island modes with a new definition of quantum numbers are predicted to have regular patterns in 
rapidly rotating stars (\citealt{Lignieres08, Lignieres09, Lignieres10, Reese08, Reese09}).
Nowadays echelle diagrams have been derived for model calculations by \citet{Deupree11} and \citet{Ouazzani15}.
 
A much higher signal to noise ratio of the space missions 
 (CoRoT -- \citealt{Baglin06, Auvergne09} and {\it Kepler} -- \citealt{Borucki10}) has yielded the 
detection of a much larger set of modes in $\delta$ Scuti stars, in some cases hundreds of modes. There is,
however, still a discrepancy between the number of actually detected and theoretically 
predicted modes. A direct comparison of the detected and model frequencies does not give a 
unique solution for mode identification. The traditional mode 
identification methods using the color amplitude ratio and the phase 
differences \citep{Garrido00, Viskum98} can only be applied to a few
of the detected frequencies and limits the advantage of space missions by 
ground-based possibilities.

Although the frequencies of the low radial order modes are out of the asymptotic 
regime so that solar-type regular frequency spacings are not expected, some underlying 
regularities in the frequency spectra have been found even from ground-based international campaigns
(\citealt{Handler97, Breger99}). In the era of great expectation of space missions, 
\citep{Dziembowski98} wondered whether mode identification will not be even more difficult to obtain 
for $\delta$ Scuti stars. \citet{Barban01} published their effort to develop an alternative 
method which does not involve any knowledge of a model in its first steps and uses very precise data, 
i.e. frequencies, rather than amplitudes and phases. Model frequencies were used with realistic 
amplitude distribution based on visibility effects. Their echelle diagram is nicely regular in the high 
frequency region ($f >$ 35 d$^{-1}$) but the echelle ridges are mixed in the low frequency region 
($f <$ 35 d$^{-1}$). Using a different $\Delta\nu$ value, some regularity appeared in the low 
frequency region, too, showing one straight and two highly inclined echelle ridges. 
After introducing the rotational splitting, the frequency spacing histogram no longer shows the peak 
at the large separation frequency, unless only the high amplitude and high frequencies are used.
 
Based on MOST data \citep{Matthews07}, CoRoT data 
(\citealt{Garcia Hernandez09, Garcia Hernandez13, Mantegazza12}) and {\it Kepler} data 
(\citealt{Breger11, Kurtz14}) successful investigations were done for individual $\delta$ Scuti stars. 
Recently \citet{Garcia15} reported investigation of frequency spacing on a sample 
of 15 {\it Kepler} $\delta$ Scuti stars, obtaining a large separation for 11 stars. Their work flow 
shows that they use the Fourier Transform (FT) to find the large separation and, in ambiguous cases,  
they make a decision based on the histogram of the frequency differences. Knowing the frequency spacing, the 
echelle diagram is derived.  In the echelle diagram for their sample case, KIC 1571717, two echelle 
ridges contain 6 and 4 frequencies, while on the other echelle ridges only three or even fewer frequencies are located.

We aimed to find a complementary method for searching for series of regular frequency spacing in a large sample  
of $\delta$ Scuti stars and for finding 
similarities or differences between the individual targets. We intended to use only the 
frequencies of high precision obtained by the space mission(s) similar to \citet{Barban01}, 
but finally we also used the amplitudes at the starting point.

\section{Search for sequences}\label{data_processing}

The motivation for searching for sequences among the frequencies is twofold. 
Neither the histogram of frequency differences nor the FT give information on the 
connections between the frequencies. Only the most frequent spacing can be 
given, although the echelle diagram plotted later with the spacing shows which 
frequencies are located on the same ridge. 
Published examples show that the highest peak of the FT sometimes agrees with half of the 
large separation and not with the large separation itself that was obtained by modeling.
On the other hand, avoiding any additional ground-based requirement for mode identification 
(nowadays we have too many targets but limited telescope time) we must know which frequencies 
are related to each other, for example, as in the sequence of eigenmodes with the same $l$ value.

\begin{figure*}
\includegraphics[width=18cm]{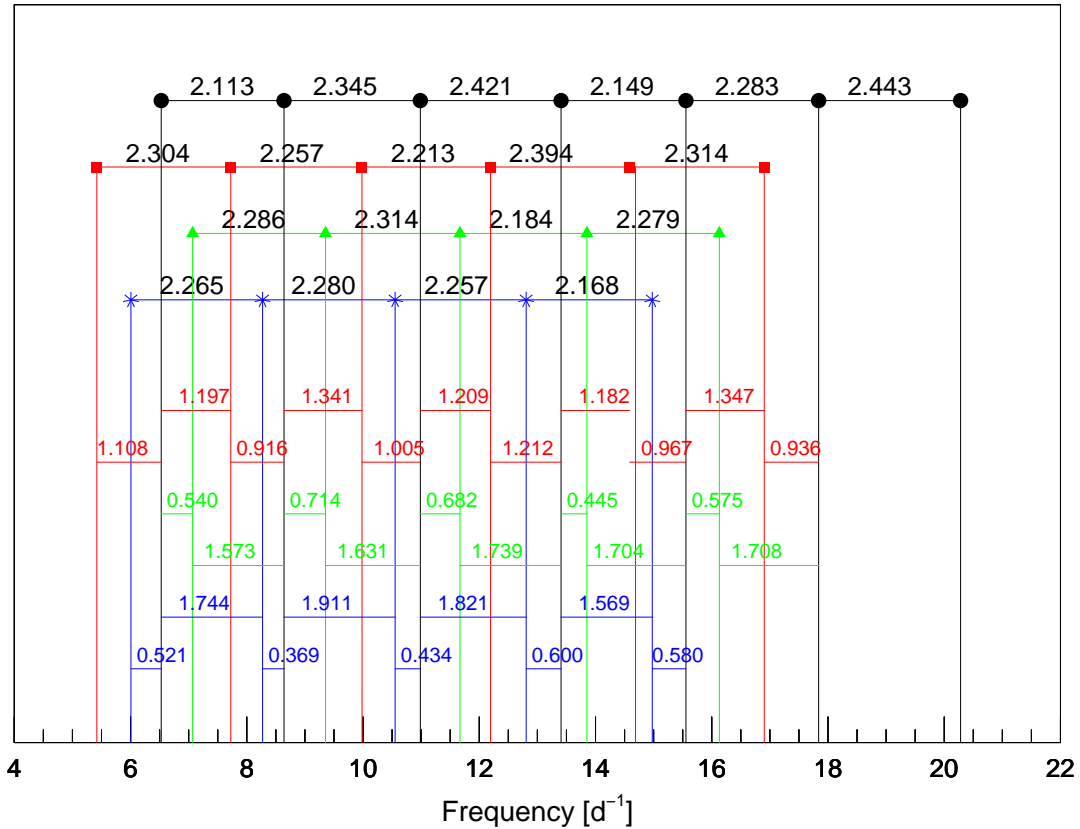}
\caption[]{
Sequences with quasi-equal spacing, and shifts of the sequences for CoRoT 102675756. 
1st -- black dots, average spacing 2.292$\pm$0.138 d$^{-1}$; 
2nd -- red squares, 2.296$\pm$0.068  d$^{-1}$; 
3rd -- green triangles, 2.265$\pm$0.057 d$^{-1}$; 
4th -- blue stars, 2.242$\pm$0.051 d$^{-1}$ average spacing was obtained. 
The mean spacing of the star is 2.277$\pm$0.088 d$^{-1}$. The shifts of the 2nd, 
3rd and 4th sequences relative to the first one are also given in the same color as the sequences.
} \label{llc_1}
\end{figure*}

A personal motivation was our result on
the $\delta$ Scuti star CoRoT 102749568 \citep{Paparo13}. The frequency 
difference of the dominant modes and the period ratio, consistent with the 
radial fundamental and first overtone, yielded a relative identification of 12 frequencies
having three different degrees, $l$. They did not show a comb-like structure, as in
the asymptotic regime, but they were regularly interwoven. 
A paper by \cite{Chen15} recently appeared on astro-ph that gave an interpretation with rotationally 
split modes, but there is no doubt about the regular spacing.

The large databases of space missions (CoRoT, {\it Kepler}) allow us great possibilities to search for any 
regularity in the frequency distribution that could help to derive the large separation or to resolve the 
interaction of rotation and pulsation in the non-asymptotic regime. 
We introduce in this paper the visual inspection and algorithmic approaches of the sequence search method.

\subsection{Visual inspection (VI)}\label{visual}

Visual inspection of the frequencies uses the ability of the human brain for searching 
for structure(s) in a seemingly unstructured sample.
Both observational \citep{Breger09} and theoretical investigations 
\citep{Suarez14} have reported that the frequency differences between 
successive radial orders are not constant, but form a quasi-periodic structure at low radial 
order. A standard deviation of such a structure is roughly 2.5 $\mu$Hz 
(0.216 d$^{-1}$) \citep{Garcia Hernandez09}, which allowed us to look 
for non-strictly equidistant structures. 

Visual inspection of the frequencies in the target stars proved to be a time-consuming but 
flexible way of searching for sequences with quasi-equidistant frequencies. 
The goal was to find criteria for an algorithmic search. 
In the asymptotic regime the theoretically predicted equidistancy in frequency (solar-type oscillations), 
or in periods (g modes) serves as a good basis, although with shortcomings \citep{vanReeth14} 
for algorithmic search \citep{Unno81, Aerts10}. For the low order p modes we do not have such definite 
guidelines given by the theory.

As a basic idea, we checked the frequency difference between the frequencies of highest amplitude pairs. 
When the pairs had a similar spacing and one member of the pair appeared in the other pair(s), 
we regarded them as a starting point of a sequence. Finding frequencies of lower amplitude with a 
similar spacing, the sequence was extended to higher and lower direction of the frequency range.
The frequency pairs that did not connect to the first sequence were used as the starting point 
for another sequence. We found 1, 2, 3 and 4 sequences for 19, 18, 17 and 10 stars, respectively. 
For 4 stars, 5 sequences were identified, while for 3 stars, 6 sequences were identified, by visual inspection.
The sequences are shifted with respect to each other.

We present here only one case.
Fig.~\ref{llc_1} shows the four sequences of similar regular spacing for CoRoT 102675756. The sequences consist of 7, 6, 5 and 5 members, 
respectively, altogether including more than 50\% of the filtered frequencies. 
In this case, each consecutive member of a sequences is excited with amplitude above the 
accepted limit (in general 0.1 mmag. The amplitude limitation was introduced to avoid 
the increasing complexity of the frequency distribution at lower amplitude levels.) 
In other cases, however, we allowed to skip one member of the 
sequence, if we did not find it, but if half of the second consecutive member's spacing matched the 
regular spacing. Fig.~\ref{llc_1} also displays the independent spacing values between 
the successive members of the sequence. The mean value of the spacing is 
independently given for each sequence in the figure caption.
The mean values differ only in the second digits. The general spacing value, 
which is the average of the sequences, is 2.277$\pm$0.088 d$^{-1}$. 
The deviation of the individual spacings from the mean value in CoRoT 102675756 and in other 
targets suggests that we may use $\pm$0.1 d$^{-1}$ tolerance in the algorithmic search. 
With this knowledge we could reduce the standard deviation of the spacing to half of the 
value given by \citet{Garcia Hernandez09}.

The shift of the sequences does not seem to be randomly distributed, but 
represents characteristic values. 
For CoRoT 102675756, we present the shifts of the appropriate members of the shifted sequences to 
the appropriate member (before and after) of the first sequence, used as a reference.
The frequencies of the second sequence are almost 
mid-way between those of the first sequence, what we expect in a comb-like structure of stars pulsating
in the asymptotic regime. 
Exactly, the averaged shift of the second sequence relative to the first is 1.024 d$^{-1}$ to the left. However,
the differences do not steadily increase when we move to the higher radial orders. The most plausible 
explanation at this level would be that we directly see the large separation, and 
the different echelle ridges have a different order, $l$, but the situation is probably
more complicated, due to the rotation. The members of the two other sequences are asymmetric, one 
of them is closer to the $n$th, 
the other to the $n+1$th member of the first sequence. We derived the shifts for our all 
targets in this concept for finding any regularity in it and comparing them to the frequency spacings. 
In our present case, however, even more regularity appears in the shifts. 
The fourth sequence is also shifted to mid-way compared to the third sequence. If we calculate 
the averaged shift of the fourth sequence relative to the third one, we get a 1.092 d$^{-1}$ shift to the left. 
The third sequence relative to the first one and the second relative to the fourth one are on average shifted to the 
right by 0.591 and 0.539 d$^{-1}$, respectively. In addition, the fourth sequence is shifted relative to the 
first one by 0.501 d$^{-1}$ to the left.

\begin{figure}
\includegraphics[width=9cm]{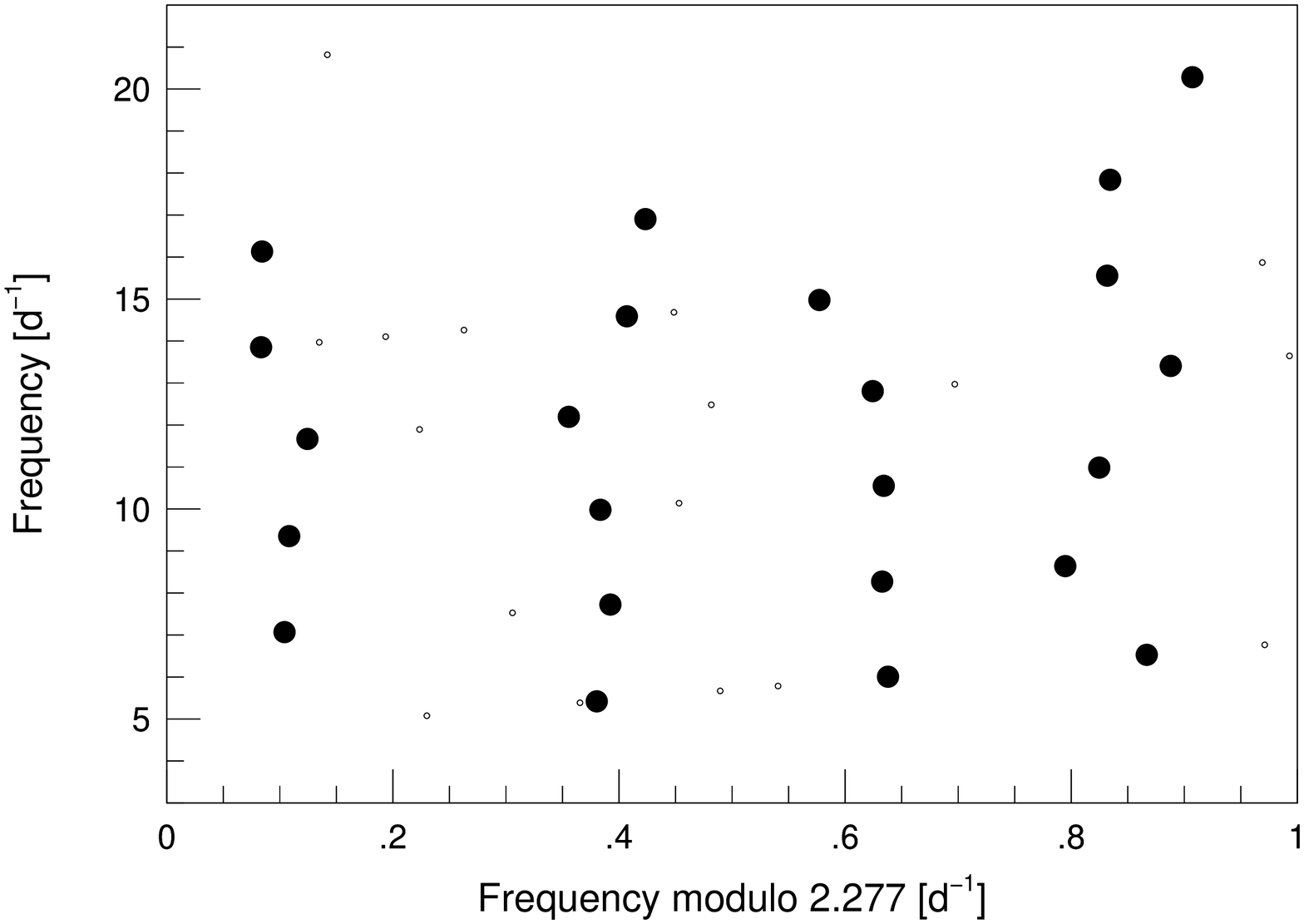}
\caption[]{
Echelle diagram of star  CoRoT 102675756, consistent with the sequences of 
Fig.~\ref{llc_1} result of the visual inspection). The mean spacing of the star was used as a modulo frequency. 
The whole frequency content of the star is plotted (small and large dots). 
The larger dots show the vertical representation of the sequences, the echelle ridges. 
} \label{llc_2}
\end{figure}

The echelle diagram (a vertical representation of the sequences) is nowadays extensively used in 
asteroseismology as a valuable way of displaying periodic spacing. The echelle diagram (frequency versus 
frequency modulo 2.277 d$^{-1}$) is presented in Fig~\ref{llc_2}. for  CoRoT 102675756, 
in agreement with the result of the visual inspection. The first, second, third and fourth sequences 
correspond to the echelle ridges at 0.85, 0.38, 0.1 and 0.65 d$^{-1}$ modulo values.  
The curvature of the echelle ridges shows the 
slightly smaller or larger value of the actual spacing between the successive 
frequencies, but it shows a high level of regularity. It is especially worthwhile to emphasize that the echelle 
ridges start from the low frequency region of the generally accepted frequency region of $\delta$ Scuti stars 
\citep{Balona11}. They cover 4-5-6 large separation regions if we interpret the spacing between the frequencies 
as the large separation. The ridge with the highest frequency is still well below the asymptotic regime of 
the p modes. We found four rather straight echelle ridges in the regime where only 1-2 echelle ridges were 
found by \citet{Barban01} and \citet{Garcia15}. Never have been so many frequencies arranged along the echelle ridges 
in a $\delta$ Scuti star as we found for CoRoT 102675756.

The small dots represent frequencies that are not located on any ridges. According to the ray 
dynamic approach of rapidly rotating stars these modes are not island modes.

According to the AAO spectral classification \citep{Guenther12, Sebastian12}, 
CoRoT 102675756 has
 $T_{\mathrm {eff}}$=7350$\pm$300 K, $\log g$=3.2$\pm$0.5 and A7III spectral type and a variable 
star classification as a $\gamma$ Dor type star \citep{Debosscher09}. 
Following the process used by \citet{Balona15} for {\it Kepler} stars, we derived a possible 
equatorial rotational velocity (100 kms$^{-1}$). 
Using the formula of \citet{Torres10} for the radius, we derived a 
first order rotational splitting (0.269 d$^{-1}$).

The three shifts of the four sequences (the second relative to 
the fourth, the fourth relative to the first and the first relative 
to the third are 0.539, 0.501 and 0.591 d$^{-1}$, respectively) could 
be interpreted as twice the value of the rotational splitting. 
The appearance of twice the value of the rotational frequency is 
predicted by theory \citep{Lignieres10}. We wonder whether the shifts 
of the second sequence relative to the first (1.024 d$^{-1}$) and the 
fourth relative to the third (1.092 d$^{-1}$) represent a higher multiple 
of the rotational frequency or are instead connected to the odd and even 
parity in the ray dynamic approach.

\begin{figure}
\includegraphics[width=9cm]{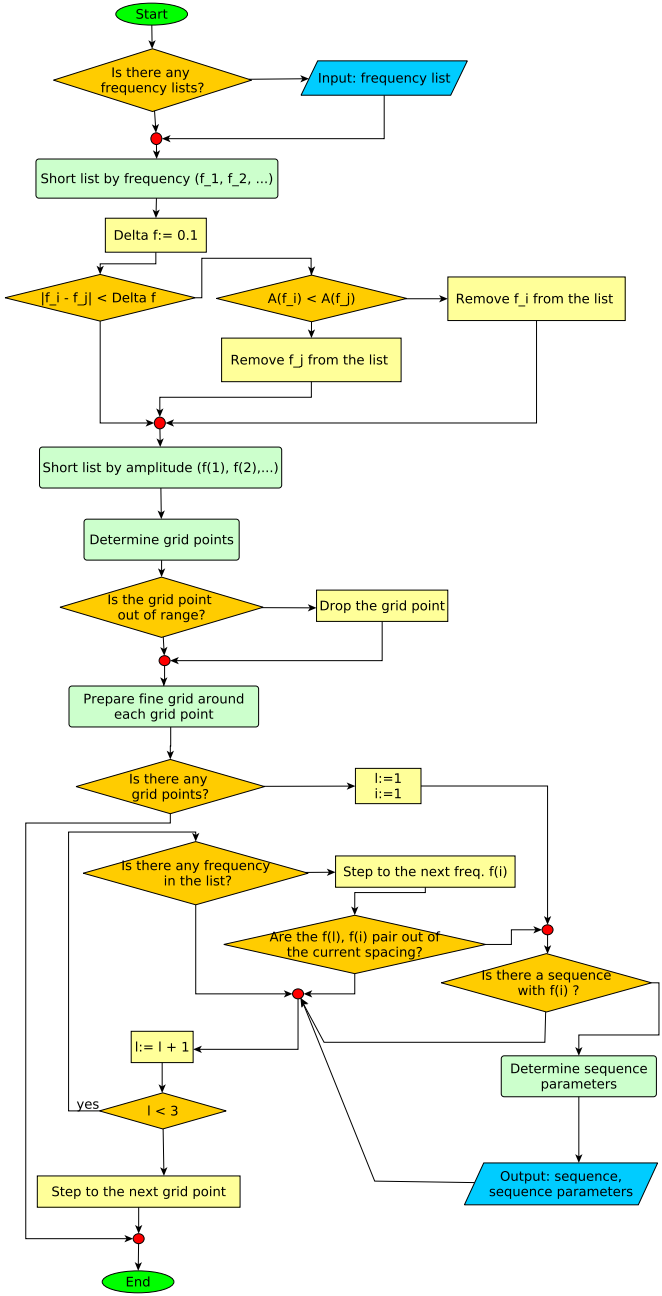}
\caption[]{
Schematic flow diagram of our Sequence Search Algorithm (SSA).
The convention in this plot is that the rightwards pointing
arrows of the conditional boxes show the `yes' destinations
and the downward pointing ones show the `no' conditions.
See the text for the details.
} \label{flow}
\end{figure}

The echelle ridges start 
at 5 d$^{-1}$, near the $\gamma$ Dor frequency range. However, these regularities appear in frequencies, and 
not in periods as we would expect for g modes of $\gamma$ Dor stars. The star could be a p-mode/g-mode hybrid like those 
discovered by space missions \citep{Uytterhoeven11}; however, \citet{Hareter13} classified it as 
a pure $\delta$ Scuti stars with no hybrid character. 

The regular structure of the shifted sequences is so obvious (at least for one of us) that already the visual 
inspection recognized sequence(s) in most of our targets. 

\begin{figure*}
\includegraphics[width=18cm]{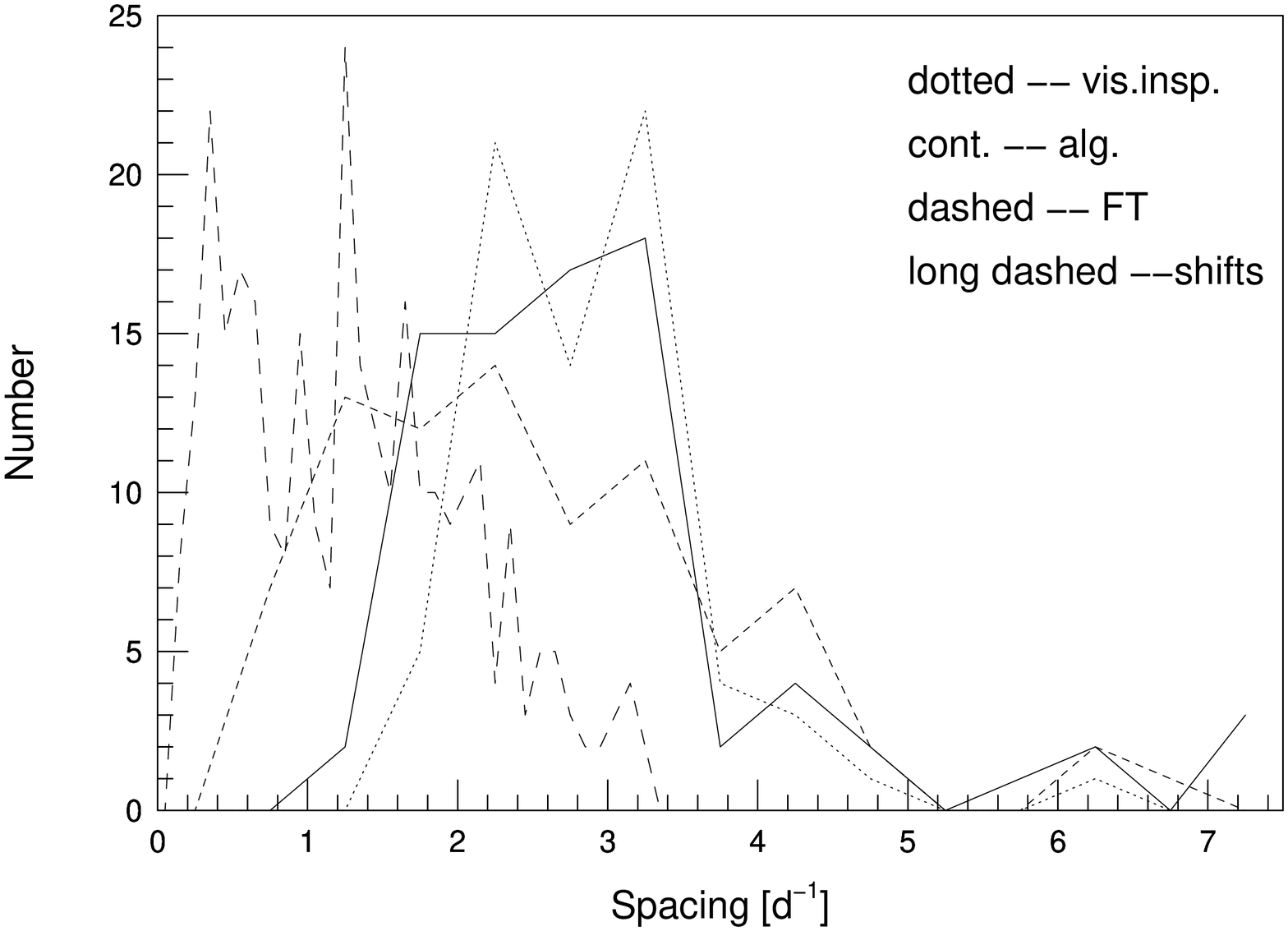}
\caption[]{
Distributions of the spacing derived by visual inspection (dotted line), algorithmic search (continuous line), 
and Fourier Transform (dashed line) are compared.
The latter one has higher numbers at low spacing values. The distribution of the shifts 
are overplotted by the large dashed line. One of the highest peaks coincides with the large 
numbers of spacings obtained by FT.
} \label{fig3}
\end{figure*}

\subsection{Algorithmic search (SSA)}

Of course, the visual inspection of a larger sample is very time-consuming.
However, detecting non-uniform period spacing can be rather complicated, especially if 
two different series with a different average spacing overlap, as \citet{vanReeth14} discuss for $\gamma$ Doradus stars.
Based on the constraints of the visual inspection (possible range of spacing, value 
of tolerance) we developed an algorithm (SSA) to automate the process. 
The main steps of the algorithm are illustrated by the schematic flow
chart in Fig.~\ref{flow}.

The SSA uses the same frequency lists as input data as we use for VI.
(1) First, if incidentally some too-close frequency pairs remained (viz. within the range 
of tolerance $\Delta f= \pm 0.1$~d$^{-1}$) in the filtered frequency lists,
we remove the lower amplitude compliments. This means few (1-5) frequencies for a couple of stars. (2) We calculate the mutual distances
between each pair for the ten highest amplitude frequencies. Then we
omit those distances which are out of the possible spacing range determined by VI (1.35-8 d$^{-1}$). (3) The
algorithm constructs a spacing grid consisting of the proper distances and fine grids
around each such distance. The fine grid allows us to find those sequences in which
the average spacing is slightly different from any of the exact
distances
between the frequency pairs within the sequence. (4) The core process of the algorithm searches for sequences at each
grid point. (i) The search starts with the
highest amplitude frequency (basis frequency) and tries to find an at
least four element frequency sequence, where the elements of the sequence are located at the
actual spacing within the value of tolerance. If the SSA find a sequence it
calculates some parameters such as the deviation of the sequence from the exact equidistant
spacing, and the sum of the amplitude of the component frequencies.
(ii) The search continues with the closest neighbor frequency to the
previously used ones, except it is farther on the basis frequency than the actual spacing.
In the first case we can find additional (shifted) sequences for the given spacing. In the latter
case we change the basis frequency to the second highest amplitude frequency and
re-start the search. According to the results of the VI we do not use the lower
amplitude frequency as the basis one. (5) At the end, we compare the parameters of the found sequences and
determine the dominant spacing, where we found the most number of sequences and/or
the most frequencies in sequences. If we have similar results for some
different spacings we take into account additional parameters (amplitude sums, standard
deviation) for choosing the dominant spacing. 

Using all frequencies in the algorithmic search, we found slightly more sequences (even 7 or 8), 
although the requirements were more severe than in the visual inspection,  but the search is more systematic.  For CoRoT 102675756 the SSA determined five sequences compared to the four sequences obtained by VI, showing that SSA does not exactly reproduce the results of the visual inspection.
In many cases SSA found more than one almost equally valid dominant spacing. For CoRoT 102675756 
both 2.249 and 1.977 d$^{-1}$ spacings were found. The difference of the two spacings (0.272 d$^{-1}$) numerically agrees with the rotation frequency (0.269 d$^{-1}$). Calculating the modulo value for the 2.249 d$^{-1}$ spacing, the echelle ridges appear at 0.14, 0.42, 0.55, 0.69 and 0.90 values, while the echelle ridges calculated with 1.977 d$^{-1}$ are at 0.03, 0.13, 0.37, 0.55 and 0.90. 
This means that four of them appear at the same modulo value, and only one ridge is different. 17 frequencies appear in ridges for both spacing but they are located on different ridges. In the echelle diagram we do not see the rotationally split frequencies as parallel ridges as is the case for the asymptotic regime for a sequence of consecutive radial orders. This suggests that none of the ridges represent frequencies with the same $l$ value and cannot be trivially interpreted as the large separation.

\subsection{Fourier Transform (FT)}\label{FT}

At the introduction of a new method it is always desirable to compare it to a previously used method.
We prepared the FT of our targets and derived the spacing as the highest peak.
We followed the way described by \cite{Handler97}. The frequency content with unit amplitudes is 
the input data for getting the spectral window of the frequency distribution. The time domain is 
transferred again to the frequency domain and the amplitudes to power. We applied the definition of the Nyquist 
frequency omitting the low-frequency region of FT.

The highest peak of the FT of CoRoT 102675756, 2.137 d$^{-1}$, is pretty much equal to the spacing of the visual 
inspection, i.e. 2.277 d$^{-1}$, or the spacing obtained by the algorithmic search: 2.249 d$^{-1}$.
However, the two methods (three approaches) do not always give as precise agreement as in the presented case. 
Mostly the visual inspection and the algorithmic search resulted in similar spacings. However, there is full agreement 
of the spacing obtained by the different approaches only for 13 targets. 
It is worthwhile to mention the 8 targets where the FT spacing proved to be half of the visual inspection's spacing. 
These stars show the closest similarity to the spacing in the asymptotic regime.

Fig.~\ref{fig3} summarizes the spacings obtained by the different approaches. 
The individual spacings derived for 77 $\delta$ Scuti stars were binned in 0.5 d$^{-1}$ wide interval. 
To avoid the overlapping of the histograms, the number in the bin is plotted at the middle of the bin and 
the points are connected. The total number under a certain type curve gives the number of stars in our sample.
The distribution of the spacings is only slightly different for the visual inspection (dotted line) and algorithmic 
search (continuous line) (1.5-3.8 d$^{-1}$) but FT (dashed line) shows a remarkably 
wider distribution (1.0-4.5 d$^{-1}$), especially in the lower spacing values. 
In these distributions we applied the spacing value resulting in echelle ridges with less scatter, although in many targets more than 
one spacing appeared, mostly around the two pronounced peaks of the visual inspection's spacing. 

We conclude that the different approaches (with different requirements) are able to 
catch different regularities among the frequencies. The different spacing values 
are not the mistake of any of the methods; rather the methods are sensitive to different regularities. 
The visual inspection and the algorithmic search concentrate on the continuous sequences, while the FT  is more sensitive to  
the number of similar frequency differences.
When we have a second sequence with a mid-shift, then the FT shows that, instead of the spacing of a single sequence,
the spacing will be twice the value of the highest peak in the FT.

If the shifts of the sequences are asymmetric, the FT shows a low and a larger value with equal probability. 
When we have many peaks in the FT then we have many echelle ridges with different shifts with respect to each other. 
The sequence method helps to explain the fine structure of the Fourier Transform.

The shifts derived between the sequences (if there is more than one) are a fraction of the spacings. 
It could be worthwhile to compare their distribution to that of the spacing.

We overplotted the distribution of the shifts in Fig.~\ref{fig3} by the long dashed line. The shifts 
are more numerous that the spacing. Their number depends on the number of the echelle ridges. 
Although we averaged them in a sequence, six shifts appears in the case of four echelle ridges as Fig.~\ref{llc_1} shows. 
The shifts are binned in 0.1 d$^{-1}$ intervals. Only the numbers in a bin versus the center frequencies of 
the bin are plotted, as in the case of the spacings.
The plot shows two dominant peaks at 0.35  and 1.25 d$^{-1}$. The latter coincides with a peak of FT spacing 
showing that in many cases the FT shows the shift of the sequence instead of the spacing of a single sequence.
 
\section {Discussion}

The basic question is why we have so many sequences in our targets. What is the origin of 
the different regularities obtained in a star? We have found many reasons for producing regularities in 
the frequencies of pulsating stars.

The alias structure as a possible source of the regularity was ruled out after detailed examination 
of the spectral window of the light curves. The other possible source, the linear combination, was 
also excluded, since only a few linear combination appeared and they did not belong to the echelle ridges. 
Only one target's single echelle ridge was excluded due to linear combination.

In a pulsating star, however, the consecutive radial orders with an $l$ value represent a 
sequence of frequencies with a regular spacing (although not exactly equidistant). 
The radial orders of different $l$ value frequencies can represent different 
sequences with the same spacing. The simplest explanation would be that the 
different echelle ridges represent sequences with different $l$ value. However, in some 
cases we have seven or even eight 
echelle ridges. Due to the geometrical cancellation, it is not very probable that we can observe 
modes with so high $l$ value. We therefore conclude that regular spacing can appear due to 
consecutive radial orders but it is definitely not the only origin.

\begin{figure}
\includegraphics[width=9cm]{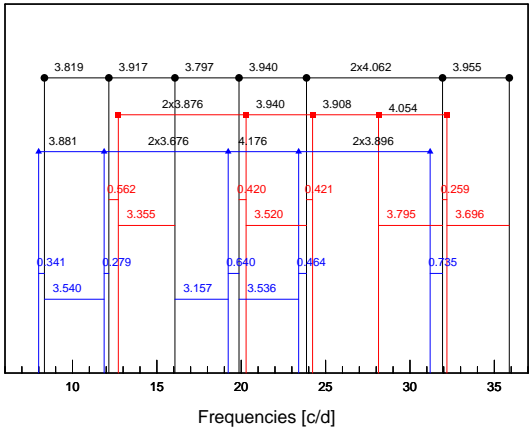}
\caption[]{
Sequences among the observed frequencies of FG Vir that is one of the
best-studied $\delta$ Scuti star. The sequences are presented in black, red and
blue colors. Some members are missing and the small shifts
of the sequences with respect to each other strengthen the feeling of grouping
of frequencies. The shifts of the members of a sequence with respect to the first
 one are plotted with the same color as the sequence.
} \label{fig14}
\end{figure}

\begin{figure}
\includegraphics[width=9cm]{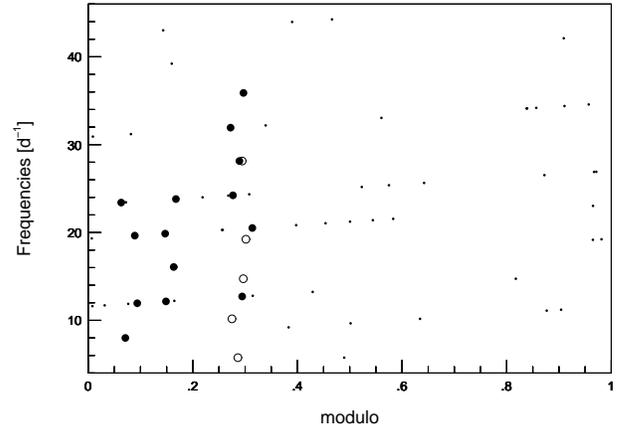}
\caption[]{
Echelle diagram of the observed frequencies of FG Vir calculated
by 3.86 d$^{-1}$ (full large dots). The background points are all the independent
frequencies. Only 21\% of the frequencies are located on the echelle ridges. Open
circles delineate the
echelle ridge calculated by the 4.47 d$^{-1}$ spacing. A side peak of FG Vir in
the FT is shown in the next figure.
} \label{fig17}
\end{figure}

\begin{figure}
\includegraphics[width=9cm]{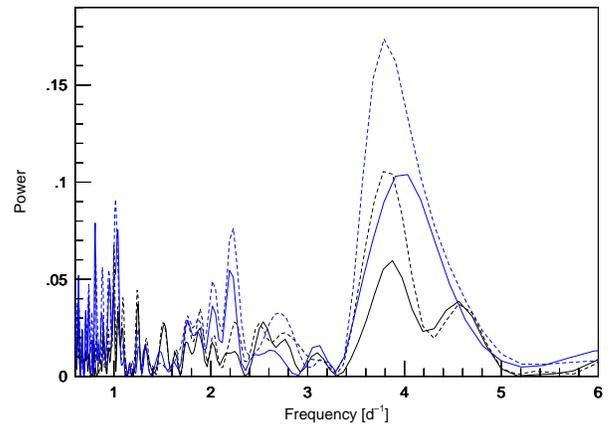}
\caption[]{
FT of the observed modes in FG Vir. We prepared five different subsets using a
color code
for compact presentation. Black continuous line -- all observed frequencies,
black dashed line -- independent frequencies, 
blue continuous line -- all frequencies above 0.4 mmag, blue dashed line -- 
independent frequencies above 0.4 mmag.
} \label{fig18}
\end{figure}

We tried to apply the theoretical period ratio of the radial fundamental and first overtone 
for the identification of a sequence with $l$=0. The appearance 
of the radial period ratio depends on the spacing value, the most popular being the 
2.1-2.9 d$^{-1}$ spacing regions with (7.1-10.24)/(9.1-13.12) frequency region 
for possible radial fundamental and first overtone period ratio. 
However, these pairs did not
 agree with the first two members of the echelle ridges, so we could not use 
them to localize radial modes.

\begin{figure*}
\includegraphics[width=18cm]{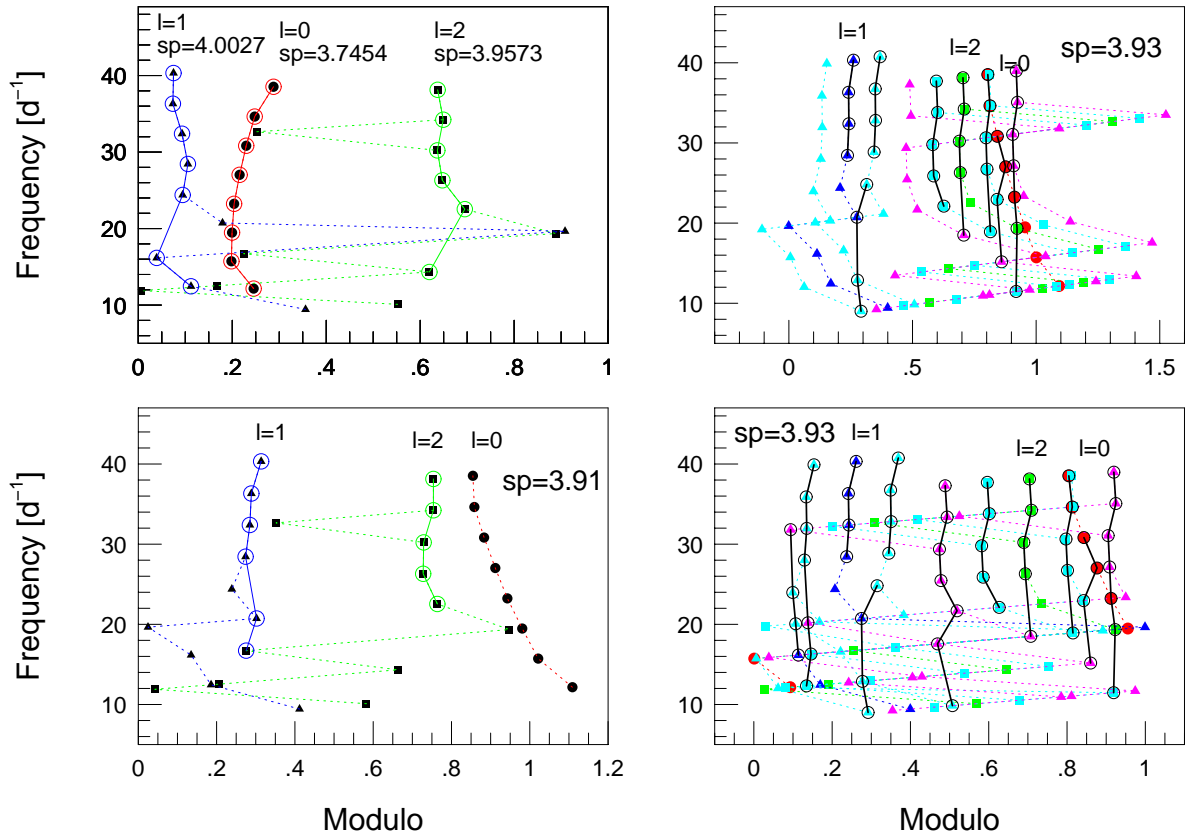}
\caption[]{
Echelle diagrams of model frequencies of FG Vir. Spacings were obtained 
independently for $l$=0, 1 and 2 modes (left, top panel) and altogether (left, bottom panel). 
The actual spacings are labeled in the panels. Right panels show the rotationally split frequencies 
in two representations: modulo values are shifted by $\pm$ 1 d$^{-1}$, if it is necessary (top panel) 
and what we see in a real star (bottom panel). Color code: red, blue and green colors are used 
for $l$=0, 1 and 2 modes, respectively. The rotationally split triplets are represented by light blue, 
while the multiplet members by magenta.
} \label{fig4}
\end{figure*}

\begin{figure*}
\includegraphics[width=18cm]{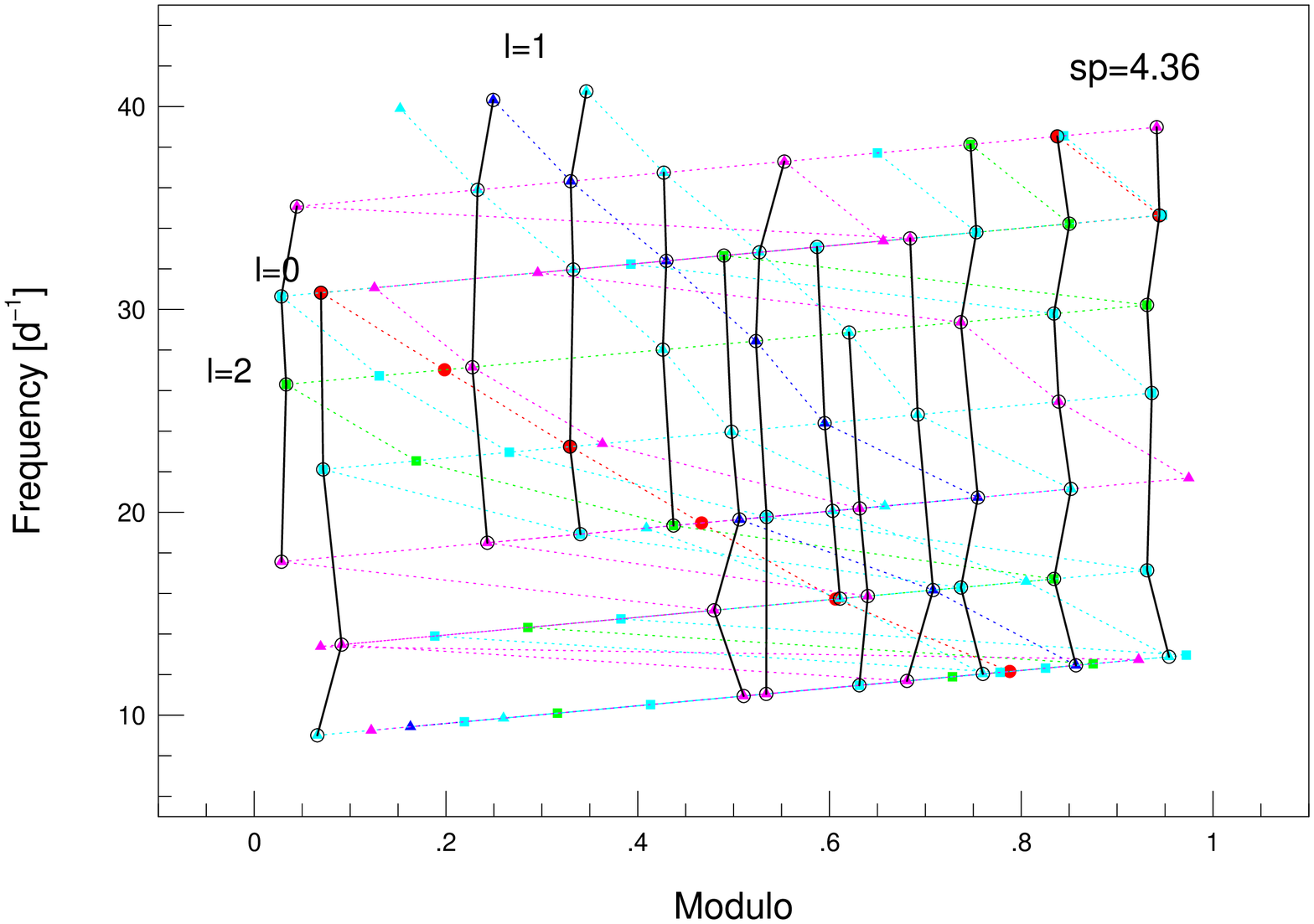}
\caption[]{
Echelle diagrams of model frequencies of FG Vir modulo 4.36 d$^{-1}$. 
The search for regular spacing resulted also in a second spacing, that we used here. 
The color code shows that the echelle ridges contain frequencies with different degrees 
$l$ and also the rotationally split frequencies too. The second spacing is the sum of the 
first spacing (3.93 d$^{-1}$) and the $f_{\mathrm{rot}}=0.4234$ d$^{-1}$. The color code is the same as in Fig. 4.
} \label{fig5}
\end{figure*}

According to the perturbative theory, the rotational splitting modifies the regular sequence of eigenmodes. 
At slow rotation the splitting value is lower and has an equidistant structure. For fast 
rotators the splitting structure is 
more complex, quintuplets are overlapped and they are not equidistant due to the second-order effects. 
The ray dynamic approach revealed different classes of modes with different characteristics 
concerning regularities. Fig. 6. of \citet{Ouazzani15} presented altogether nine echelle ridges (three for the non-rotating case and six for the rotating case) 
for $\hat{l}$ = 0, 1 and 2 island modes (which are the counterparts of the low order acoustic modes) 
with odd and even parity.
With the odd and even parity the number of the echelle ridges is doubled.
Some of the echelle ridges are overcrossed. Due to the curvature in the low and high frequency 
region and the overcrossing region, our tolerance limit could not recognize and resolve all echelle ridges in their model.

Of course, a final solution could be a pulsational modeling of a real star with their ACOR code \citep{Ouazzani12}. 
It seems that we are close to this level of interpreting intermediate and fast rotating stars. However, first, 
we checked our sequence search method for the first order rotationally split modes of the 
best observed $\delta$ Scuti star, FG Vir.

\subsection {FG Vir as a check star}

A trivial check is the application of our sequence search method 
for the observed frequencies of a well-studied $\delta$ Scuti star and 
for $l=0,1,2$ eigenmodes of its modeling.  
FG Vir is one of the best studied $\delta$ Scuti stars from both the observational (photometry and spectroscopy) 
and from the theoretical side. Our sequence search method (both the visual inspection and algorithmic search) was applied for 75 observed frequencies were taken from \cite{Breger05}. 
The sequences, using the independent frequencies, were prepared by visual inspection. Fig.~\ref{fig14} shows 17 frequencies in three 
sequences
with 3.931$\pm$0.122 d$^{-1}$ mean spacing. The second and the third sequences are 
slightly shifted with respect to
the first one. Some of possible members of the sequences are not excited over
the amplitude limit.
It is quite understandable that FG Vir was one example of a $\delta$ Scuti star
showing groups of frequencies.
The shifts of the second and third sequences relative to the first one on 
average are 0.414 and 0.491 d$^{-1}$, respectively. Concerning the 0.423 d$^{-1}$ 
rotational frequency of FG Vir obtained by \citet{Mantegazza94}, this shift should 
be interpreted as the rotational splitting of triplets. The triplets introduce a
spacing that is the sum of the large separation and the rotational splitting.
Regarding the prograde or retrograde component of a triplet, or the components of
a multiplet structure,
another possible spacing is the sum of the large separation and twice the value of
 the rotational splitting.

The algorithmic search revealed 3.86 d$^{-1}$ spacing for the observed frequencies in 
good agreement with the 3.7 d$^{-1}$ spacing published by \cite{Breger09}.
Fig.~\ref{fig17} shows the echelle diagram with 3.86 d$^{
-1}$ spacing obtained by SSA. The sporadic distribution (not more than three 
members of a possible echelle ridge) does not fulfill our requirements for SSA. 
Three echelle ridges with 14 frequencies (full dots) are shown. Open circles show an 
overplotted echelle ridge that
was calculated using a 4.47 d$^{-1}$ spacing, which is the second spacing obtained by
 SSA.  This spacing represents the sum of the large separation and the rotational frequency,
although it has a slightly larger value than the 0.432 d$^{-1}$ rotational
frequency obtained by \citet{Mantegazza94}.

We also applied the FT for the observed
frequencies. We checked two effects, the
linear combination frequencies and the effect of the lower amplitude frequencies,
for the FT.
Fig.~\ref{fig18} gives a compact representation of our results. We prepared the
FT for the
following data sets: all the observed frequencies of FG Vir (black continuous
line), only the independent
frequencies (black dashed line), all frequencies above 0.4 mmag
amplitude (48 frequencies, blue continuous line) and the independent frequencies
 above 0.4 mmag
amplitude (42 frequencies, blue dashed line).
The dominant features of the curves are the dominant peaks at around 3.9 d$^{-1}
$ in all subsets.
Concerning the dominant peaks we can conclude that the
black lines (both continuous and dashed) reflect a second lower peak at 4.56 d$^
{-1}$, in addition to the dominant spacing.
Comparing this value to the second spacing of FG Vir obtained by SSA (4.47 d$^{-1}$), we may interpret
this second spacing as the sum of the large separation and the rotational frequency. Using only
 the high amplitude modes
(blue lines) we miss the side peak
of the FTs, but the dominant peaks have a larger width covering the second peaks
 of the FT. It seems to
be plausible that the rotationally split modes have a lower amplitude. The other
 effect is that the subsets
using only the independent modes have a dominant peak with a higher amplitude.

\begin{figure}
\includegraphics[width=9cm]{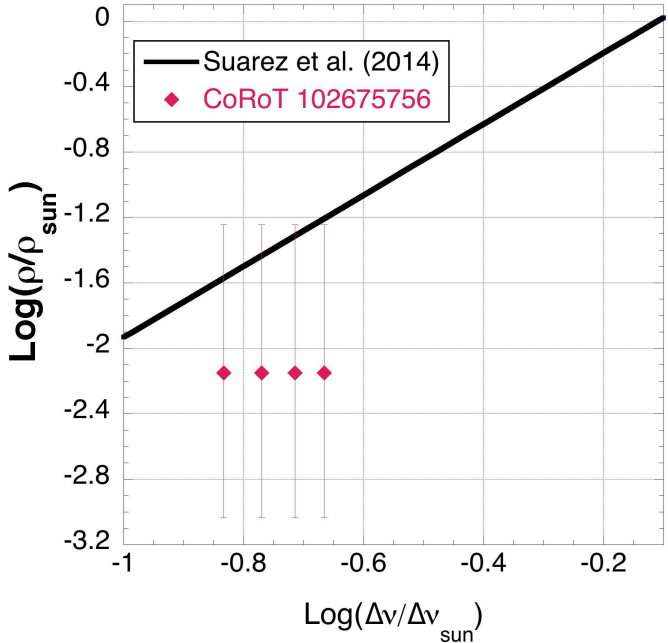}
\caption[]{
The location of the four possible large separation of CoRoT 102675756 on the space density versus large separation diagram given by \citet{Suarez14}. The error bars are calculated based on the error bars of $T_{\mathrm {eff}}$ and $\log g$. The $\Delta\nu$=1.710 d$^{-1}$ is the closest point to the straight line. 
} \label{fig6}
\end{figure}

The FG Vir model that we used here (see \citealt{Guzik00} for modeling details) 
was evolved using the OPAL opacities with \cite{Grevesse93}
abundance mixture, initial helium mass fraction Y=0.28, and mass
fraction of elements heavier than hydrogen and helium, Z, of 0.02.  It has
$T_{\mathrm{eff}}=7419$~K, $L=13.92$~L$_{\sun}$, and $R$=2.26~R$_{\sun}$.

This model has a main-sequence age of 0.867 Gyr, and a convective core with
core hydrogen mass fraction 0.270.  Outside the convective core there is a
composition gradient where a high Brunt-V\"ais\"al\"a frequency has developed
that forms a cavity in which the mode frequency is less than both the
Brunt-V\"ais\"al\"a frequency and the Lamb frequency, and where gravity modes can
propagate.

For the $l=1$ modes, the ten calculated frequencies range from pure p modes
with 8 radial nodes at highest frequency, to pure g modes (2 g-type nodes)
at lowest frequency.  For $l=2$, the twelve calculated frequencies range from
pure p modes with 7 radial nodes to pure g modes (4 g-type nodes).  At
intermediate frequencies, the modes may have a mixture of p and g-type
nodes.  When the g-mode character of the modes begins to dominate, nearly
equidistant frequency spacing is altered.  For $l=1$, this occurs for the
lowest frequency modes.   For $l=2$, this occurs both for the lowest
frequencies when the g-mode character dominates, and for the third highest
frequency when a g-type node first appears.   For $l=0$, since these are
radial modes, the modes cannot have a nonradial g-mode character, and the
calculated frequency spacing follows a regular pattern.

The $l$=0 , $l$=1 and $l$=2 eigenmodes were independently treated,
resulting 3.7454, 4.0027 and 3.9573~d$^{-1}$ spacing, respectively. The corresponding echelle 
ridges are shown in the left, top panel of Fig.~\ref{fig4}. The $l$=0 eigenmodes have a slightly 
smaller spacing than the $l$=1 and 2 modes. Applying the method for all eigenmodes, 
we found a 3.91 d$^{-1}$ spacing and only two echelle ridges ($l$=1 and 2), presented in the left,
 bottom panel of Fig.~\ref{fig4}. Due to 
the smaller spacing value of $l$=0 frequencies, the curvature of the $l=0$ frequencies according 
to modulo 3.91 d$^{-1}$, did not fulfill the tolerance requirement of SSA.

The test shows that the consecutive radial orders of an $l$ value can represent echelle ridges in our 
sequence method. However, not all of the eigenmodes (connected by dotted lines) are situated on the 
echelle ridges and, due to the slightly different spacing of the eigenmodes, we do not find all $l$ ridges 
at the same time.

The first order rotational splitting of $l=1$ modes to triplet and $l=2$ modes
to a quintuplet structure was derived using a reliable $f_{\mathrm{rot}}$=0.4234 d$^{-1}$, obtained from observation.
 
Application of the method for the 94 rotationally split modes leads to a more conclusive result for 
the unexpectedly large number of echelle ridges in our sample. 
We found two dominant spacings, at 3.93 d$^{-1}$ which is similar in the observed frequencies. 
The other spacing is 4.36 d$^{-1}$ which is the sum of the previous one and the  $f_{\mathrm{rot}}$. 
Since we know how the rotationally split modes were generated, we show two representations for the 
3.93 d$^{-1}$ spacing on the right side of Fig.~\ref{fig4}. To follow the $l$=0, 1 and 2 
eigenmodes and the rotational splitting clearly, the modulo values were shifted by $\pm$1 as necessary (top panel). 
The bottom right panel gives the situation that we can see in a real star. The echelle diagram 
modulo 4.36 d$^{-1}$ spacing is given in Fig.~\ref{fig5}. It is obvious that in all cases we see many 
echelle ridges (8, 11 and 13). 
The color code shows that the echelle ridge contains not only the consecutive radial orders of 
an $l$ value, but the rotationally split modes, too. The severe constraint of the tolerance shows that 
the complex echelle ridges have quasi-equal spacing despite their origin.

We conclude that one spacing reflects the large separation, while the other 
spacing represents the sum of the large separation and the rotation. We found a good argument 
for the numerous echelle ridges in $\delta$ Scuti stars, 
 even using first order rotational splitting. The echelle ridges do not overcross but we have 
closely spaced echelle ridges. However, in this case we can resolve them with our $\pm$0.1 d$^{-1}$ tolerance level.

What may we conclude for CoRoT 102675756 given the results on the model frequencies of FG Vir? Although the difference of the two spacings for CoRoT 102675756 (2.249 and 1.977 d$^{-1}$) agrees with the estimated rotational splitting  
(0.269 d$^{-1}$) we may not be sure that one of the spacings reflects the large separation, 
itself, or instead a combination of the large separation and the rotational frequency.
Taking into account that the appearance of 
the rotational frequency and/or double its value have been found in the auto-correlation of the 
frequency spectrum \citep{Lignieres10}, 
we have the following possibilities as an explanation: 
\begin{eqnarray}
SP_1 & =& \Delta\nu, \ {\mathrm {and}} \ SP_2 = \Delta\nu - \Omega_{\mathrm{rot}}, \nonumber \\
SP_2 & =& \Delta\nu,  \ {\mathrm {and}} \ SP_1 = \Delta\nu + \Omega_{\mathrm{rot}}, \nonumber \\
SP_1 & =& \Delta\nu + 2\cdot\Omega_{\mathrm{rot}}, \ {\mathrm {and}} \ SP_2 = \Delta\nu + \Omega_{\mathrm{rot}}, \nonumber \\
SP_2 & =& \Delta\nu - 2\cdot\Omega_{\mathrm{rot}}, \ {\mathrm {and}} \ SP_1 = \Delta\nu - \Omega_{\mathrm{rot}}, \nonumber
\end{eqnarray}
where, $SP_1$ and $SP_2$ are the larger and smaller values of the spacings, respectively, 
found by SSA, $\Delta\nu$ is the large separation in the traditionally used sense,
namely, the differences between the consecutive radial orders with
the same spherical degree,
and $\Omega_{\mathrm{rot}}$ is the estimated rotational frequency.

In all cases $SP_1-SP_2 = \Omega_{\mathrm{rot}}$, the estimated rotational frequency of 
CoRoT 102675756. Using $SP_1 = 2.249$ d$^{-1}$, $SP_1 = 1.977$ d$^{-1}$
and $\Omega_{\mathrm{rot}}$ = 0.269 d$^{-1}$, we get four possible value for the 
large separation ($\Delta\nu$) of CoRoT 101675756,
namely 2.249, 1.977, 1.710 or 2.517 d$^{-1}$.

Fig.~\ref{fig4} shows how these possible large separations are related to the 
relation between the mean density and the large separation given by \citet{Suarez14}. 
We obtained the mean density of CoRoT 102675756 using the formulas of \citet{Torres10} 
for mass and radius, given $T_{\mathrm{eff}}$ and $\log g$ and their uncertainties, 
and assuming solar metallicity.  Since we do not have special single-star oriented spectroscopy, 
the error bars are rather large. Nevertheless, the $\Delta\nu$=1.710 d$^{-1}$ is located 
at the closest place to the relation. This means that the larger spacing value gives the sum of 
the large separation and twice the value of the rotational frequency, while the smaller 
spacing represent the sum of the large separation and the rotational frequency.
 
\section{Conclusion}

None of the three approaches, visual inspection, algorithmic search or Fourier Transform, can give a unique spacing 
for many stars. In the simplest cases (one or two ridges) the 
spacing values are the same for all methods. When we have a complex spacing 
structure, then it may happen that any two of the three methods agrees, but 
sometimes all of them produce a different spacing. This seemingly contradictory 
result means that the methods (with different requirements) are more sensitive 
to one of the characteristic spacings. 

The benefit of the sequence search method is that, beside obtaining the spacing 
value(s), we immediately know how many sequences are found in the star. The shifts 
between sequences provide more insight into the pulsation-rotation interaction, especially 
when the shifts agree with the rotational frequency or twice its value.  
High quality spectra providing accurate physical parameters 
($\log g$, $T_{\mathrm {eff}}$, metallicity, and rotation rate) would be enough to 
interpret the spacings and obtain the seismic parameters of any large sample of $\delta$ Scuti stars.
 
The application of the sequence search method for the model frequencies of FG Vir 
was especially informative. For the rotationally split set of frequencies, 
the sequence search method revealed many echelle ridges showing that frequencies of 
different origin (consecutive radial orders or rotationally split frequencies) are located 
on the same echelle ridges with the same spacing. The two spacings found, 
3.93 and 4.36 d$^{-1}$, proved to be the large separation and the sum of the large separation 
and the rotational frequency. For our sample target, CoRoT 102675756, 
both spacings (2.249 and 1.977 d$^{-1}$) seem to be the combination of the large separation 
and the rotational frequency. The explanation that 
$2.249 = \Delta\nu + 2\cdot\Omega_{\mathrm{rot}}$ and $1.977 = \Delta\nu + \Omega_{\mathrm{rot}}$ 
seems to be more plausible, resulting in a 1.710 d$^{-1}$ large separation,  because it fits 
better the \citet{Suarez14} relation.

Our whole sample will be published in more detail in \citet{Paparo15}. 
We have good hope to make progress in 
the resolution of rotation and pulsation, reaching the asteroseismology level, too, in the non-asymptotic regime.

%\

\acknowledgments{
This work was supported by the grant: ESA PECS No 4000103541/11/NL/KLM. 
}


\begin{thebibliography}{}

\bibitem[Aerts et al.(2010)]{Aerts10}
Aerts, C., Christensen-Dalsgaard, J., \& Kurtz, D. W.\ 2010, Asteroseismology, A\&A
 Library, (Berlin - Heidelberg: Springer)
\bibitem[Auvergne et al.(2009)]{Auvergne09}
Auvergne, M., Bodin, P., Boisnard, L., et al.\ 2009, \aap, 506, 411
\bibitem[Baglin et al.(2006)]{Baglin06}
Baglin, A., Auvergne, M., Barge, P., et al.\ 2006,
in ESA SP 1306, The CoRoT Mission Pre-Launch Status - Stellar Seismology and Planet Finding,
 ed. M. Fridlund, A. Baglin, J. Lockhard, \& L. Conroy, (Noordwijk: ESA), 33
\bibitem[Balona \& Dziembowski(2011)]{Balona11} 
Balona, L. A., \& Dziembowski, W. A.\ 2011, \mnras, 417, 591
\bibitem[Balona et al.(2015)]{Balona15} 
Balona, L. A., Daszynska-Daszkiewicz, J. \& Pamyatnykh, A.A. \ 2015, \mnras, 437, 1476 
\bibitem[Barban et al.(2001)]{Barban01} 
Barban, C., Goupil, M.-J., Van't Veer-Menneret, C., \& Garrido, R.\ 2001, in ESA SP 464, 
Proc. SOHO 10/GONG 2000 Workshop, ed. A. Wilson, (Noordwijk: ESA), 399
\bibitem[Borucki et al.(2010)]{Borucki10} 
Borucki, W. J., Koch, D., Basri, G., et al.\ 2010, Science, 327, 977
\bibitem[Breger et al.(1999)]{Breger99} 
Breger, M., Pamyatnykh, A. A., Pikall, H., \& Garrido, R.\ 1999, \aap, 341, 151
\bibitem[Breger et al.(2005)]{Breger05} 
Breger, M., Lenz, P., Antoci, V., et al.\ 2005, \aap, 435, 955
\bibitem[Breger et al.(2009)]{Breger09} 
Breger, M., Lenz, P., \& Pamyatnykh, A. A.\ 2009, \mnras, 396, 291
\bibitem[Breger et al.(2011)]{Breger11} 
Breger, M., Balona, L., Lenz, P., et al.\ 2011, \mnras, 414, 1721
\bibitem[Chen \&  Li(2015)]{Chen15} 
Chen, X. H., \& Li, Y.\ 2015, arXiv:1508.03916
\bibitem[Deupree(2011)]{Deupree11} 
Deupree, R. G.\ 2011, \apj, 742, 9
\bibitem[Debosscher et al.(2009)]{Debosscher09} 
Debosscher, J., Sarro, L. M., L{\'o}pez, M., et al.\ 2009, \aap, 506, 519
\bibitem[Dziembowski et al.(1998)]{Dziembowski98} 
Dziembowski, W. A., Balona, L. A., Goupil, M.-J., \& Pamyatnykh, A. A.\ 1998, in The First MONS Workshop: Science
with a Small Space Telescope, ed. H. Kjeldsen, \& T. R. Bedding, ({\AA}rhus: {\AA}rhus Univ.), 127
\bibitem[Fox Machado et al.(2006)]{Fox06} 
Fox Machado, L., P\'erez Hern\'andez, F., Su\'arez, J.C., et al.\ 2009, \aap, 446, 611
\bibitem[Garc\'{\i}a Hern\'andez et al.(2009)]{Garcia Hernandez09} 
Garc\'{\i}a Hern\'andez, A., Moya, A., Michel, E. et al.\ 2009, \aap, 506, 79
\bibitem[Garc\'{\i}a Hern\'andez et al.(2013)]{Garcia Hernandez13} 
Garc\'{\i}a Hern\'andez, A., Moya, A., Michel, E. et al.\ 2013, \aap, 559, A63
\bibitem[Garc\'{\i}a Hern\'andez et al.(2015)]{Garcia15} 
Garc\'{\i}a Hern\'andez, A., Ligni\`eres, F., Balona, L., et al.\ 2015, in EPJ Web of Conf. 101, 
The Space Photometry Revolution, ed. R. A. Garc\'{\i}a, \& J. Ballot, id.06026
\bibitem[Garrido(2000)]{Garrido00} 
Garrido, R.\ 2000, in ASP Conf. Ser. 210, Delta Scuti and Related Stars, ed. M. Breger \& M. H. Montgomery, 
 (San Francisco, CA: ASP), 67
\bibitem[Goupil et al.(2005)]{Goupil05} 
Goupil, M.-J., Dupret, M. A., Samadi, R., et al.\ 2005, J. Astrophys. Astron., 26, 249
\bibitem[Grevesse \& Noels(1993)]{Grevesse93} 
Grevesse, N., \& Noels, A. 1993, Physica Scripta, T47, 133
\bibitem[Guenther et al.(2012)]{Guenther12} 
Guenther, E. W., Gandolfi, D., Sebastian, D., et al.\ 2012, \aap, 543, A125
\bibitem[Guzik et al.(2000)]{Guzik00} 
Guzik, J. A., Bradley, P. A., \& Templeton, M. R..\ 2000, in ASP Conf. Ser. 210,
Delta Scuti and Related Stars, ed. M. Breger \& M. H. Montgomery, (San Francisco, CA: ASP), 247
\bibitem[Handler et al.(1997)]{Handler97} 
Handler, G., Pikall, H., O'Donoughue, D., et al.\ 1997, \mnras, 286, 303
\bibitem[Hareter(2013)]{Hareter13}
Hareter, M.\ 2013, PhD Thesis, Univ. Vienna, Austria
\bibitem[Jackson et al.(2004)]{Jackson04} 
Jackson, S., McGregor, K.B., \& Skumanich, A.\ 2004, \apj, 606, 1196
\bibitem[Kurtz et al.(2014)]{Kurtz14} 
Kurtz, D. W., Saio, H., Takata, M., et al.\ 2014, \mnras, 444, 102
\bibitem[Ligni\`eres et al.(2006)]{Lignieres06} 
Ligni\`eres, F., Rieutord, M., \& Reese, D. R.\ 2006, \aap, 455, 607
\bibitem[Ligni\`eres et al.(2008)]{Lignieres08} 
Ligni\`eres, F., \& Georgeot, B.\ 2008, \pre, 78, 6215
\bibitem[Ligni\`eres et al.(2009)]{Lignieres09} 
Ligni\`eres, F., \& Georgeot, B.\ 2009, \aap, 500, 1173
\bibitem[Ligni\`eres et al.(2010)]{Lignieres10} 
Ligni\`eres, F., Georgeot, B., \& Ballot, J.\ 2010, Astron. Nachr., 331, 1053
\bibitem[Mantegazza et al.(1994)]{Mantegazza94}
Mantegazza, L., Poretti, E., \& Bossi, M.\ 1994, \aap, 287, 95
\bibitem[Mantegazza et al.(2012)]{Mantegazza12} 
Mantegazza, L., Poretti, E., Michel, E., et al.\ 2012, \aap, 542, A24
\bibitem[Matthews(2007)]{Matthews07} 
Matthews, J. M.\ 2007, CoAst, 150, 330
\bibitem[McGregor et al.(2007)]{McGregor07} 
McGregor, K. B., Jackson, S., Skumanich, A., \& Metcalf, T. S.\ 2007, \apj, 663, 560
\bibitem[Ouazzani et al.(2012)]{Ouazzani12} 
Ouazzani, R.-M., Dupret, M.-A., \& Reese, D. R.\ 2012, \aap, 547, A75
\bibitem[Ouazzani et al.(2015)]{Ouazzani15} 
Ouazzani, R.-M., Roxburgh, I. W., \& Dupret, M.-A.\ 2015, \aap, 579, A116
\bibitem[Papar\'o et al.(2013)]{Paparo13} 
Papar\'o, M., Bogn\'ar, Zs., Benk\H{o}, J. M., et al.\ 2013, \aap, 557, A27
\bibitem[Papar\'o et al.(2015)]{Paparo15}
Papar\'o, M., Benk\H{o}, J. M., Hareter, M., \& Guzik, J. A.\ 2015, \apjs \ (submitted)
\bibitem[Reese et al.(2006)]{Reese06} 
Reese, D. R., Ligni\`eres, F., \& Rieutord, M.\ 2006, \aap, 455, 621
\bibitem[Reese et al.(2008)]{Reese08} 
Reese, D. R., Ligni\`eres, F., \& Rieutord, M.\ 2008, \aap, 481, 449
\bibitem[Reese et al.(2009)]{Reese09} 
Reese, D. R., Thompson, M. J., MacGregor, K. B., et al.\ 2009, \aap, 506, 183
\bibitem[Roxburgh (2006)]{Roxburgh06} 
Roxburgh, I. W.\ 2006, \aap, 454, 883
\bibitem[Saio et al.(2015)]{Saio15} 
Saio, H., Kurtz, D. W., Takata, M., et al.\ 2015, \mnras, 447, 3264
\bibitem[Sebastian et al.(2012)]{Sebastian12} 
Sebastian, D., Guenther, E. W., Schaffenroth, V., et al.\ 2012, \aap, 541, A34
\bibitem[Su\'arez et al.(2010)]{Suarez10} 
Su\'arez, J. C., Goupil, M.-J., Reese, D. R., et al.\ 2010, \apj, 721, 537
\bibitem[Su\'arez et al.(2014)]{Suarez14} 
Su\'arez, J. C., Garc\'{\i}a Hern\'andez, A., Moya, A., et al.\ 2014, \aap, 563, A7
\bibitem[Torres et al.(2010)]{Torres10} 
Torres, G., Andersen, J. \& Gimen\'ez, A. 2010, \aapr, 18, 67
\bibitem[Unno et al.(1981)]{Unno81} 
Unno, W., Osaki, Y., Ando, H, et al.\ 1981, Nonradial Oscillations of Stars, 
(Tokyo: Univ. of Tokyo Press), 2nd ed.
\bibitem[Uytterhoeven et al.(2011)]{Uytterhoeven11}
Uytterhoeven, K., Moya, A. Grigahc\`ene, A., et al.\ 2011, \aap, 534, A125
\bibitem[Van Reeth et al.(2014)]{vanReeth14} 
Van Reeth, T., Tkachenko, A., Aerts, C., et al.\ 2014, \aap, 574, A17
\bibitem[Viskum et al.(1998)]{Viskum98} 
Viskum, M., Kjeldsen, H., Bedding, T., et al.\ 1998, \aap, 335, 549
\bibitem[Zahn(1992)]{Zahn92} 
Zahn, J.-P.\ 1992, \aap, 265, 115

\end{thebibliography}
\end{document}